\documentclass[twocolumn,showpacs,floatfix,pre,superscriptaddress,numbers,sort&compress]{revtex4-1}
\RequirePackage{lineno}

\usepackage{graphicx} 
\usepackage{dcolumn} 
\usepackage{epsf} 
\usepackage{amsmath}
\usepackage{calrsfs}
\usepackage{bm} 
\usepackage{setspace}
\usepackage{color}
\usepackage[toc,page]{appendix}
\usepackage{natbib}

\begin{document}

\title{Structure of the simple harmonic-repulsive system in liquid and glassy states studied by the triple correlation function}

\author{V.A.~Levashov}
\affiliation{Technological Design Institute of Scientific Instrument Engineering, 630055, Novosibirsk, Russia. E-mail: valentin.a.levashov@gmail.com}
\affiliation{Institute for High Pressure Physics, Russian Academy of Sciences,108840, Moscow (Troitsk), Russia}

\author{R.E.~Ryltsev}
\affiliation{Institute for High Pressure Physics, Russian Academy of Sciences,108840, Moscow (Troitsk), Russia}
\affiliation{Institute of Metallurgy, UB RAS, 620016, 101 Amundsen str., Ekaterinburg, Russia}
\affiliation{Ural Federal University, 620002, 19 Mira str,, Ekaterinburg, Russia}

\author{N.M.~Chtchelkatchev}
\affiliation{Institute for High Pressure Physics, Russian Academy of Sciences,108840, Moscow (Troitsk), Russia}
\affiliation{Moscow Institute of Physics and Technology,141700, Institutskiy per.9, Dolgoprudny, Moscow Region, Russia}
\affiliation{Ural Federal University, 620002, 19 Mira str,, Ekaterinburg, Russia}


\begin{abstract}
An efficient description of the structures of liquids and, in particular, 
the structural changes that happen with liquids on supercooling remains to be a challenge. 
The systems composed of soft particles are especially interesting in this context because 
they often demonstrate non-trivial local orders that do not allow to introduce the concept of the nearest-neighbor shell.
For this reason, the use of some methods, developed for the structure analysis of atomic liquids, is questionable for the soft-particle systems.
Here we report about our investigations of the structure of the simple harmonic-repulsive liquid in 3D using the 
triple correlation function (TCF), i.e., the method that does not rely on the nearest neighbor concept.
The liquid is considered at reduced pressure ($P=1.8$) at which it exhibits remarkable stability against crystallization on cooling.
It is demonstrated that the TCF allows addressing the development of the orientational correlations 
in the structures that do not allow drawing definite conclusions from the studies of the bond-orientational order parameters.
Our results demonstrate that the orientational correlations, if measured by the heights of the peaks in the TCF, 
significantly increase on cooling.
This rise in the orientational ordering is not captured properly by the Kirkwood's superposition approximation.
Detailed considerations of the peaks' shapes in the TCF suggest the existence of a link between the orientational 
ordering and the slowdown of the system’s dynamics.
Our findings support the view  that the development of the orientational correlations in liquids may play a significant role in the liquids' 
dynamics and that the considerations of the pair distribution function may not be sufficient to understand 
intuitively all the structural changes that happen with liquids on supercooling.
In general, our results demonstrate that the considerations of the TCF are useful 
in the discussions of the liquid's structures beyond the pair density function 
and interpreting the results obtained with the bond-orientational order parameters.
\end{abstract}



\today
\maketitle


\section{Introduction}\label{s:intro}

Despite many years of research, 
understanding of the structure of liquids and glasses remains to be 
a challenge~\cite{Tanaka2019,wei2019assessing,Ryu2019SciRep,Wang2019ActaMater,Wang2019JALCOM,Tong2019NatComm,Royall2015,EMa2011}.
Of course, there are many methods for addressing liquids' structures~\cite{Tanaka2019,wei2019assessing,Royall2015,EMa2011,Stukowski_2012}. 
Among these methods, the method of the pair distribution function (PDF) represents the method of first resort. 
However, the complete understanding of liquids requires knowledge of the whole spectrum of the multiparticle 
correlations~\cite{Bogoliubov1946,born1946general,Kirkwood1946}.
In general, the approaches based on considerations of the many-body correlation functions are well 
known~\cite{hansen1990theory,bogoljubov1960problems,bogoliubov1962problems,born1946general,yvon1935theorie,Tanaka2020}.

In particular, many studies of the triple correlation function (TCF) 
have been made~\cite{kirkwood1935statistical,alder1964triplet,egelstaff1971experimental,raveche1972three,
bhatia1976triplet,gubbins1978thermodynamic,haymet1981triplet,haymet1985orientational, stillinger1988theoretical,
muller1993triplet,ZahnK2003,Ru2003,grouba2004superposition,vaulina2004three,coslovich2011locally,coslovich2013static,ho2006three,
singh2014triplet,donko2017higher,galimzyanov2017three,dhabal2017probing,stillinger2019structural}. 
However, investigations of the TCF are significantly less common than the considerations of the PDF. 
One reason for this situation is the complexity of the TCF, which depends on three arguments, while the PDF depends on one argument only. 
Another reason is that it is challenging to extract information about 
the TCF experimentally \cite{hansen1990theory,egelstaff1971experimental,gubbins1978thermodynamic}. 
Of course, this is not a problem for computer simulations.

It has been argued that the development of the angular or orientational correlations in liquids is particularly 
important for understanding the dynamics of supercooled liquids in the proximity 
of the glass transition \cite{Tanaka2019,Royall2015,EMa2011,Tanaka2020,sciortino2001debye}.
In this context, the TCF is of special interest because 
it allows addressing radial and angular (orientational) correlations at the same time.

The most often, the orientational ordering in liquids is addressed with 
the bond-orientational order (BOO) parameters 
(BOOPs)~\cite{steinhardt1983bond,hansen1990theory,Stukowski_2012,EMa2011,lechner2008accurate,walter2013booshortcoming}.
On the other hand, in our view, it is not always easy to interpret the results obtained with the BOOPs.
The reason for this situation, in our view, is related to the nature of the BOOPs, i.e., 
they are integral characteristics of the whole nearest-neighbor shells which are already relatively complex objects. 
Thus, besides considering the BOOPs, it might be reasonable to consider 
also the orientational ordering through simpler objects, for example, through the TCF.
Another important point is that the BOOPs give information only about 
the short-range order of the first coordination shell. 
However, a proper description of the medium-range order (MRO) 
may be crucially important when studying supercooled liquids and 
glasses~\cite{Mauro2011PRB,Nakamura2015Nanotech,Wu2015NatureComm,Trady2017JNCS,Guo2019JALCOM,hallett2018local}.

In the last three decades there has been increasing interest in modeling of the particle systems like colloids, polymers, macromolecules, micelles, e.t.c.
\cite{Gokhale2016,Kleman20031,Lang20001,Louis20001,Likos20011,Likos20012,Likos20021,Likos20061,Malescio20071,
Malescio20081,Frenkel20091,Prestipino20091,Saija20091,Malescio20111,Berthier20101,Zamponi2011,LuZY20111,
Mohanty20141,Xu20141,Xu20151,Xu20181,Denton20161,Bagchi20181,Cipelletti2018,Fomin2008,Ryltsev2013,Ryltsev2013PRE,Ryltsev2015,Ryltsev2017,Komarov2018SoftMatt,Komarov2019JCP}.
The interactions between the particles in such systems are quite different from the typically 
considered interatomic interactions~\cite{Lang20001,Likos20021,Komarov2018SoftMatt,Komarov2019JCP}.
In particular, the adopted interactions are much softer than the interatomic interactions.
Moreover, the interaction potentials can be finite even at zero separation between the particles
\cite{Lang20001,Louis20001,Likos20012,Likos20021}.
One interaction potential that has been used for the modeling of the soft-matter systems is the harmonic-repulsive
potential \cite{Berthier20101,Zamponi2011,LuZY20111,Xu20141,Xu20151,LuZY20111,Xu20181,
Lang20001,Likos20011,Frenkel20091,Prestipino20091,Saija20091,Zamponi2011,Denton20161,Levashov20161,Lokshin20181}.

One structural feature of many systems interacting through soft potentials is the splitting of the first peak in 
the PDF ~\cite{levashov2019anomalous,Fomin2010PRE}.
While this splitting is quite common in the soft-matter systems, 
it is unusual for one-component systems with strong repulsion at short distances. 
For example, such systems as hard spheres and metallic glasses.

The well-expressed first peak in the PDFs of the systems with strong repulsion 
led to the appearance of the concept of the nearest neighbor shell and the concept of the nearest neighbor.
In systems with strong repulsion, the interaction with second 
neighbors usually is much weaker than the interaction with the first neighbors.
Several methods for the description of the structures of 
the systems with strong repulsion based on the 1st neighbor concept, 
for example, the common neighbor analysis (CNA) and the BOO analysis.
We note that the Voronoi's construction analysis does not rely on 
on the concept of the nearest neighbor.
Since in many soft matter systems the first peak exhibits splitting, 
it is necessary to recognize that the concept of the nearest neighbor becomes ambiguous. 
In such systems, usually, there is nonnegligible interaction with the second 
neighbors (note, however, that the second neighbors are also ambiguously defined in such systems).
Therefore, in the analysis of the structures of the soft-matter systems, 
it might be reasonable to address them with methods that do not rely on 
the concept of the nearest neighbors. 
One such method is the method of the TCF.

In our previous publication, we, in particular, 
investigated the structure of the harmonic-repulsive liquid at
selected densities using the BOOPS~\cite{levashov2019anomalous}. 
From the obtained results, in our view, it is difficult to understand 
intuitively the strengths of the orientational correlations and their dependence on the temperature. 
Therefore, we reconsider the studied system using the TCF.

It is important to note, in consideration of the TCF, 
that there are exact relations, known as the Bogoliubov-Born-Green-Kirkwood-Yvon (BBGKY) hierarchy,
which relate lower-order correlation functions to higher-order correlation functions
\cite{hansen1990theory,bogoljubov1960problems,bogoliubov1962problems,born1946general,yvon1935theorie}.
One of these relations is usually used to calculate the PDFs.
Thus, if for the simple system the interaction potential and the TCF are known, then then the PDF can be found.
However, if the TCF is not known, it is still possible to proceed in a heuristic way, using
artificially introduced closure relation(s) (a guess) that expresses the TCF in terms of the PDF.
The simplest of these closure relations is the Kirkwood's superposition approximation (KSA) \cite{kirkwood1935statistical}
which states that the properly defined TCF, $g_3(r_{12},r_{13},r_{23})$, and the PDF, $g_2(r_{12})$, are related via:
\begin{eqnarray}
g_3(r_{12},r_{13},r_{23}) = g_2(r_{12})g_2(r_{13})g_2(r_{23}).
\label{eq:KSA}
\end{eqnarray}

In general, to achieve good agreement between the ``experimental" and calculated PDFs different closure relations
are needed for different systems \cite{hansen1990theory}.
However, none of the artificially introduced closure relations leads to the perfect agreement between
the PDFs calculated via the exact integral equations and the PDFs obtained in simulations \cite{hansen1990theory,raveche1972three,bhatia1976triplet,haymet1981triplet,haymet1985orientational,stillinger1988theoretical,muller1993triplet,ZahnK2003,Ru2003,grouba2004superposition}.
On the other hand, thus calculated PDFs capture many features of the simulated PDFs
and they also capture general features of the evolution of the PDFs with the temperature.

From the perspective of the present work, an important point, in our view,
is that none of the closure relations, besides the KSA can be considered as truly intuitive.
Yet, it is important to note that the KSA is not the best closure approximation (in fact, it is, probably, the worst).
For example, one obvious shortcoming of the KSA is that it follows from it that every peak 
in the PDF leads to the existence of the corresponding equilateral triangle--often this is 
not the case \cite{alder1964triplet,stillinger1988theoretical,stillinger2019structural}.
Consider, for example, the BCC lattice.
Some triangles with other geometries ``predicted" by the KSA also may not exist in
reality \cite{alder1964triplet,stillinger1988theoretical,stillinger2019structural}.

To summarize, we assume that considering the TCF extracted directly from 
the simulated trajectories may be especially useful when studying the following systems: 
(1) systems with non-trivial short-range order like soft-particles systems 
for which standard methods of short-range order analysis are not quite applicable; 
(2) supercooled liquids and glasses in which 
the medium-range orientational order may play an important role. 
Here we consider a system for which both of the mentioned cases realize. 
Thus, we use the TCF for addressing the structure of 
the simple harmonic-repulsive system in the liquid and glassy states.
In particular, we found that the orientational correlations, 
expressed by the heights of the TCF peaks, significantly 
increase on the approach of the glass transition. 
We found even more significant correlations between 
the widths of some TCF peaks and the glass transition. 
Thus, in the vicinity of the equilibrium breakdown temperature, 
the widths of some TCF peaks experience jumps.

The paper is organized as follows. In section \ref{sec:model}
we describe the model system and provide the details of our simulation procedure.
In section \ref{sec:triple} we describe the our implementation of the triple correlation function.
In section \ref{sec:results} we describe our results. We conclude in section \ref{sec:conclusion}

\section{The model and details of the simulation procedure \label{sec:model}}

The interaction between the pairs of particles in the studied model is described by the harmonic-repulsive pair potential:
\begin{equation}
  u(r)  =
  \begin{cases}
    \epsilon \left(1-\frac{r}{\sigma}\right)^2, & \text{if $r \leq \sigma$} \\
     0, & \text{if $r > \sigma $} \\
  \end{cases}\label{eq:hrp}
\end{equation}
In our simulations and the description of our results further in the paper,
we measure energy in the units of $\epsilon$,
distance in the units of $\sigma$,
and time in the units of $\tau = \left(m\sigma^2/\epsilon\right)^{1/2}$.

We used the LAMMPS molecular dynamics package to generate liquids' structures at
different pressures and temperatures \cite{Plimpton1995,lammps}.
The Nose-Hoover non-Hamiltonian equations have been used to generate the coordinates
and velocities of particles (via the ``npt" and ``iso" commands within the LAMMPS).

Practically all results reported in this paper have been obtained on the system containing 8000 particles.
Some of the obtained results were compared with the results obtained on the system consisting of 65000 particles.
From these comparisons, which we do not discuss here, we concluded that there are essentially
no size effects in the results which we discuss in this paper.

The used value of the time step at $T>0.010$ was $\delta t = 0.001\tau$,
while at $T<0.010$ the used value of the time-step was $\delta t = 0.010\tau$.
For $T<0.010$ the used value of the Nose-Hoover time-parameter used for
the temperature equilibration within the LAMMPS was $1\tau$, i.e., 100 time steps,
while the used value of the time-parameter for the pressure equilibration was $10\tau$,  i.e., 1000 time steps.
These are the recommended values for these parameters \cite{lammps}.

Initially, we generated the system as the FCC lattice at a very low density of  $\rho=0.04$.
Then, the system was melted and equilibrated at $T=0.015$.
After the equilibration (which happens very fast at $T=0.015$),
the system has been cooled at $P=0.020$ down to $T=0.010$ which is still above any observable crystallization temperature for this system.
Then, at $T=0.010$, we increased the pressure from $P=0.020$ to $P=1.8$.
Then we equilibrated the system at $T=0.010$.
At this high temperature, the equilibration time is smaller than $100 \tau$,
as can be judged from the dependence of the potential energy on time.

Then, we cooled the system.
The typical cooling rate used in our simulations was $10^6$ time steps per $\Delta T = 0.001$.
It follows from our considerations of the dependence of the mean square displacement on temperature, that,
at the discussed conditions, the equilibration time is smaller than $10^6$ MD steps, i.e.,
the time used for the equilibration of the systems.
Then, at every temperature and pressure, 100 structures have been saved with the time-interval of $10^5$ steps.
It follows from the analysis of the data that
thus produced configurations are more than sufficient for our purposes here.

The analysis of the generated structures, in all discussed cases,
has been made with the self-made programs.

\section{The triple correlation functions}\label{sec:triple}
\begin{figure}
\begin{center}
\includegraphics[angle=0,width=2.0in]{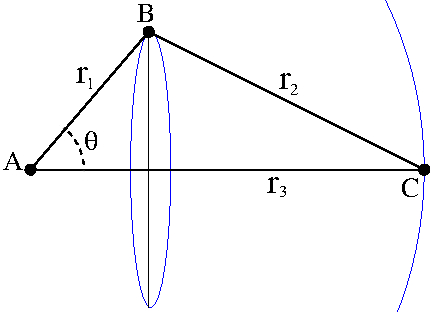}
\caption{
An image for the definition of the triple correlation function (TCF).
See Eq.~\ref{eq:triple1}.
}\label{fig:trianglefig}
\end{center}
\end{figure}
For the convenience of further considerations, it is useful to introduce 
the triple correlation function (TCF) as follows.
Let us first consider the case of randomly distributed particles.
Let particle ``$A$" be one of the vertexes of the triangle ``ABC", 
as shown in Fig. \ref{fig:trianglefig}.
It is easy to see that the number of triangles, with respect 
to a given particle ``A", with the sides of lengths $(r_1,r_2,r_3)$ in the random case is given by:
\begin{eqnarray}
dN = 4\pi r_3^2 dr_3\rho_o \cdot 2\pi r_1\sin(\theta)r_1 d\theta dr_1 \rho_o,
\label{eq:triple1}
\end{eqnarray}
where $\rho_o$ is the average density of the particles.
It follows from the law of cosines for the triangles that for the fixed $r_1$ and $r_3$
we have: $\sin(\theta) d\theta = (r_2 dr_2)/(r_1 r_3)$.
Therefore (\ref{eq:triple1}) can be rewritten as:
\begin{eqnarray}
dN = 8\pi^2 \rho_o^2 r_1 r_2 r_3 dr_1 dr_2 dr_3\;.
\label{eq:triple2}
\end{eqnarray}
Let us assume that $\rho_o = 1/a^3$, where $a$ is the average separation between the particles.
Then we define:
\begin{eqnarray}
\tilde{r}\equiv r/a\;\;\;\;\;\;\textrm{and}\;\;\;\;\;\; g_2(\tilde{r})\equiv \rho(r)/\rho_o.
\label{eq:grdef}
\end{eqnarray}
With these definitions we rewrite (\ref{eq:triple2}) as:
\begin{eqnarray}
dN = 8\pi^2 \tilde{r}_1 \tilde{r}_2 \tilde{r}_3 d\tilde{r}_1 d\tilde{r}_2 d\tilde{r}_3.
\label{eq:triple2scaled}
\end{eqnarray}

Let us consider now a non-random distribution of particles characterized by the pair density function $\rho(r)$.
One simple way to generalize expression (\ref{eq:triple2}) for this case is to assume that:
\begin{eqnarray}
dN = 8\pi^2 \left(\rho_o^2 r_1 r_2 r_3 dr_1 dr_2 dr_3\right) \left[\frac{\rho(r_1)\rho(r_2)\rho(r_3)}{\rho_o^3}\right].
\label{eq:triple3}
\end{eqnarray}
Expression (\ref{eq:triple3}) essentially represents the Kirkwood's superposition
approximation (KSA) \cite{kirkwood1935statistical,hansen1990theory,grouba2004superposition}.
Using the reduced parameters (\ref{eq:grdef}), we rewrite (\ref{eq:triple3}) as:
\begin{eqnarray}
dN = 8\pi^2 \left( \tilde{r}_1 \tilde{r}_2 \tilde{r}_3 d\tilde{r}_1 d\tilde{r}_2 d\tilde{r}_3\right) \left[g_2(\tilde{r}_1)g_2(\tilde{r}_2)g_2(\tilde{r}_3)\right].
\label{eq:triple4}
\end{eqnarray}

It is possible, instead of (\ref{eq:triple3}), to use some other approximations.
In the following, however, we will consider only expression (\ref{eq:triple3}), i.e., the KSA.

It follows from (\ref{eq:triple4}) that it is reasonable to define the triple correlation function (TCF) in the following way:
\begin{eqnarray}
g_3(\tilde{r}_1, \tilde{r}_2, \tilde{r}_2) \equiv \frac{dN}{8\pi^2 \tilde{r}_1 \tilde{r}_2 \tilde{r}_3 d\tilde{r}_1 d\tilde{r}_2 d\tilde{r}_3},
\label{eq:tcfdef}
\end{eqnarray}
where $dN$ is the number of triangles, involving a chosen particle, with the side lengths
in the intervals $(\tilde{r}_1,\tilde{r}_1 + d\tilde{r}_1)$, $(\tilde{r}_2,\tilde{r}_2+ d\tilde{r}_2)$,
and $(\tilde{r}_3,\tilde{r}_3+d\tilde{r}_3)$.

With definition (\ref{eq:tcfdef}), within the KSA, we should have, according to (\ref{eq:triple4}):
\begin{eqnarray}
g_3(\tilde{r}_1,\tilde{r}_2,\tilde{r}_3) = g_2(\tilde{r}_1)g_2(\tilde{r}_2)g_2(\tilde{r}_3).
\label{eq:ksa1}
\end{eqnarray}

In our further considerations, we will always measure distances in reduced units, $\tilde{r}$.
Therefore, in the following, for the briefness of the notations, we will omit upper tildes everywhere.

\section{Analysis of the obtained data}\label{sec:results}

Previously, it has been demonstrated that the system of particles interacting through
the harmonic-repulsive pair potential crystallizes into several different crystal
structures at different pressures~\cite{Levashov20161,LuZY20111}.
In this paper, we discuss the results only at one particular pressure, i.e., at P=1.8.
At this pressure, the studied system exhibits remarkable stability against crystallization, i.e.,
we did not observe crystallization of the this system even in very long cooling runs~\cite{levashov2019anomalous}.

\subsection{The dependence on temperature of the potential energy, volume of the system, and the
diffusion coefficient}

In this subsection, we briefly describe how the potential energy, the volume of the system, and the diffusion coefficient
depend on temperature (at $P=1.8$). These dependencies provide a general insight into the relevant temperature scales.

\begin{figure}
\begin{center}
\includegraphics[angle=0,width=3.5in]{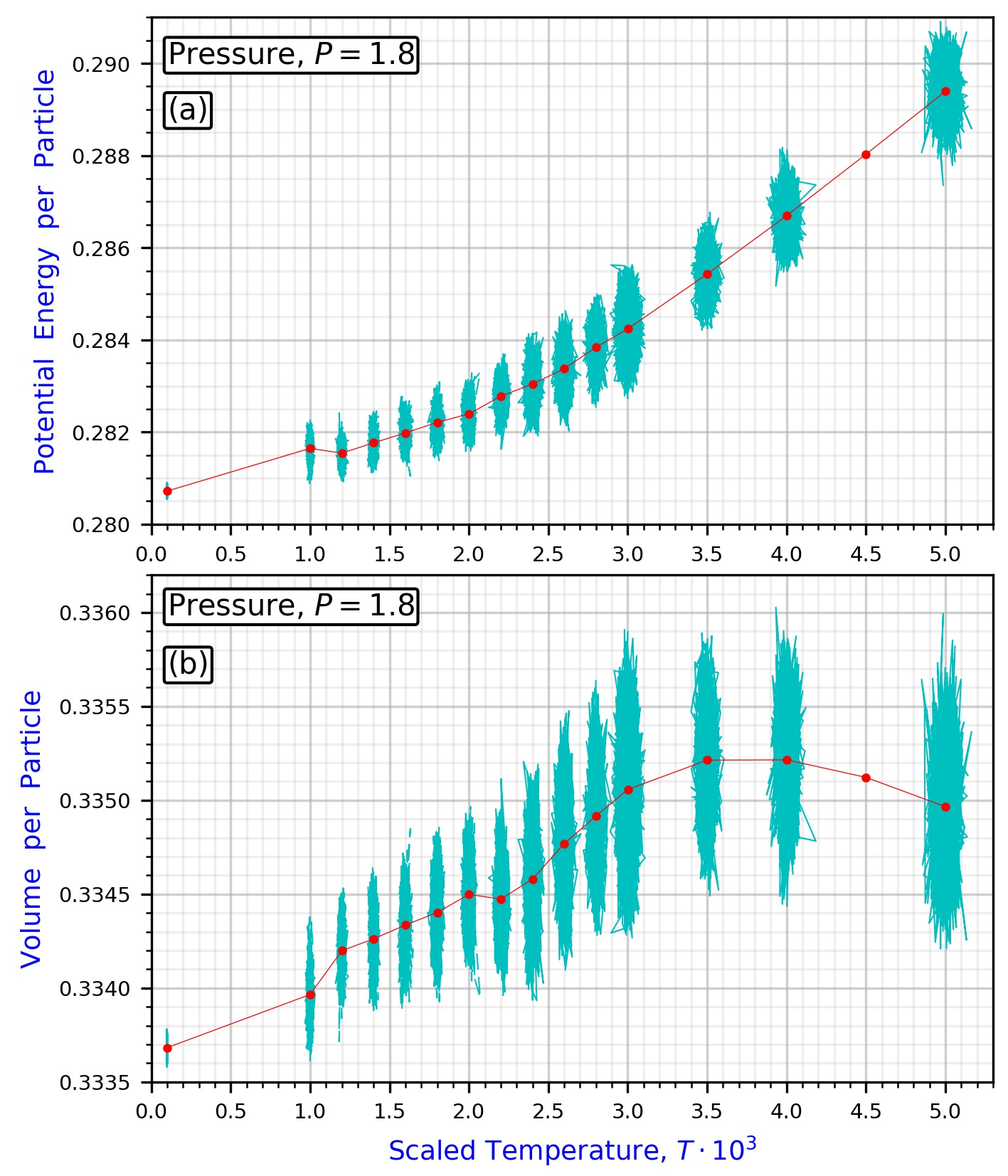}
\caption{
(a). The dependence of the potential energy per particle on temperature, $T$, at pressure $P=1.8$.
(b). The dependence of the volume of the system per particle on the temperature at $P=1.8$.
Note that at $10^3 T >3.5$ the volume of the system per particle behaves in an anomalous way,
i.e., the volume of the system decreases as the temperature increases.
}\label{fig:PE-and-V-P-1x8}
\end{center}
\end{figure}

Panel (a) of Fig. \ref{fig:PE-and-V-P-1x8} shows how the potential energy of the system depends on temperature.
The teal blobs show the results obtained directly from the NPT simulations at a given target pressure and temperatures.
The red dots show the results of the averaging of the simulation data for every set of the target parameters $(T,P)$.
From the data presented further, it follows that the glass transition in the system occurs in the temperature interval between
$T=0.0020$ and $T=0.0030$. However, the potential energy curve does not exhibit any special behavior in this interval.
We see that as the temperature is reduced from $T=0.0050$ to $T=0.0001$ the potential energy per particle
changes by $\Delta U/U \approx [(2.893 - 2.807)/2.893] \approx 0.03$.

Panel (b) of Fig. \ref{fig:PE-and-V-P-1x8} shows the dependence of the volume of the system per particle on temperature.
As it has been discussed already in Ref. \cite{levashov2019anomalous},
in the interval of temperatures $T \in [0.0035:0.0050$
the system exhibits an anomalous behavior, i.e., negative thermal expansion coefficient.
The behavior of the curve exhibits a clear change in the region of temperatures
$T \in [0.0020:0.0025]$, i.e., in the interval closely corresponding to the glass transition temperature.

\begin{figure}
\begin{center}
\includegraphics[angle=0,width=3.5in]{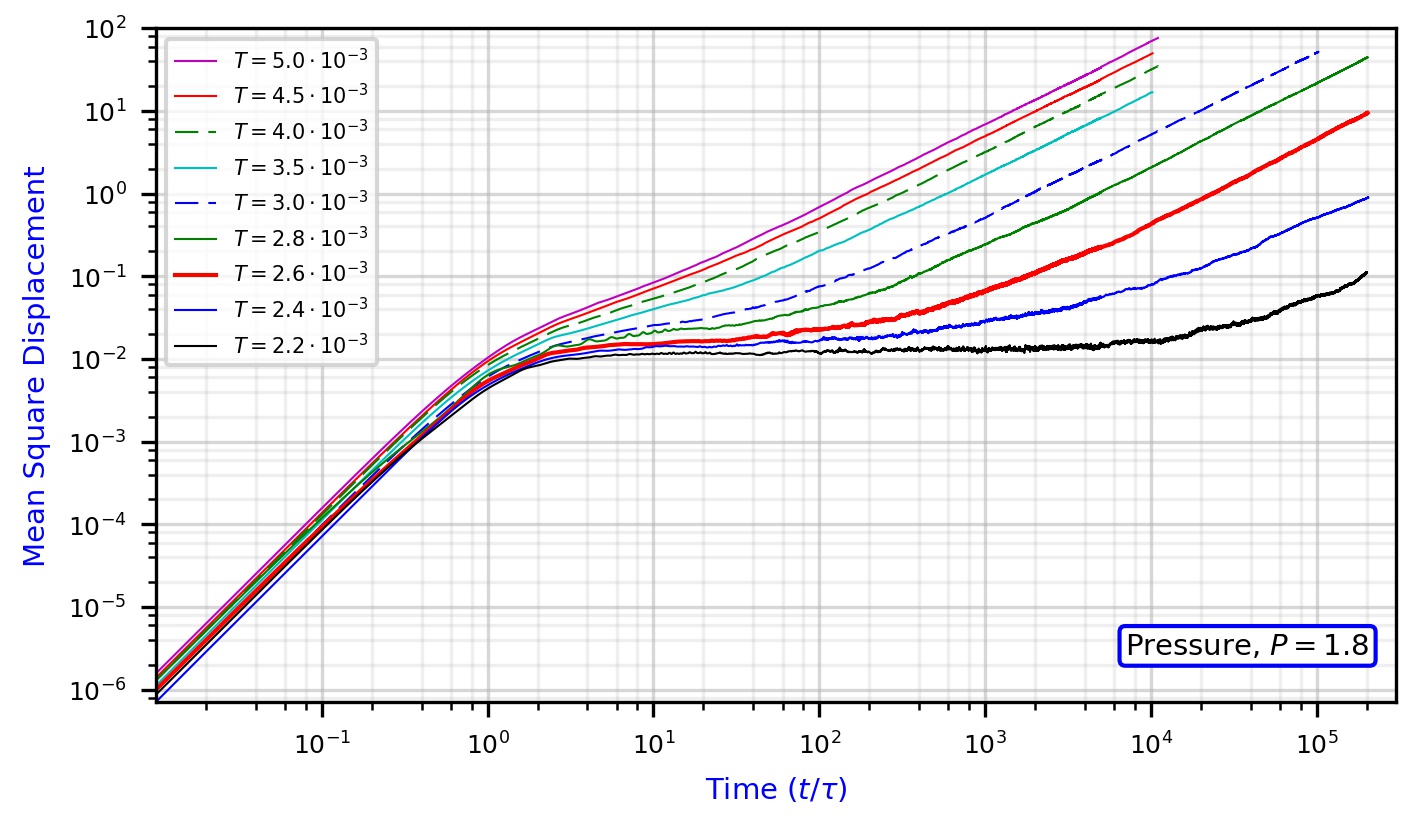}
\caption{
The dependencies of the mean square particle displacement on time
at selected temperatures at $P=1.8$.
According to the figure, we further assume that
the glass transition temperature lies in the interval
$[2.0 \leq 10^3 T_g \leq 2.6]$.
}\label{fig:meansquare}
\end{center}
\end{figure}

In Fig. \ref{fig:meansquare} we show the dependence of the mean square particle displacement on time
at selected temperatures. It follows from the figure that significant slow down of the dynamics
happens in the interval of temperatures $T \in [0.0020:0.0026]$.
In our simulations, we determined the reliable values of the long-time diffusion coefficient
at temperatures $T \geq 0.0026$. As it also can be seen from Fig. \ref{fig:meansquare},
at $T=0.0024$ and $T=0.0022$ we did not achieve the reliable values of the long-time diffusion
coefficient in simulations runs of lengths $10 \cdot 10^6$ and $20 \cdot 10^6$ MD steps.
As our goal in this publication is to study the behavior of the TCF,
the accurate determination of the diffusion coefficient at $T < 0.0026$ is beyond the scope of this paper.

\begin{figure}
\begin{center}
\includegraphics[angle=0,width=3.5in]{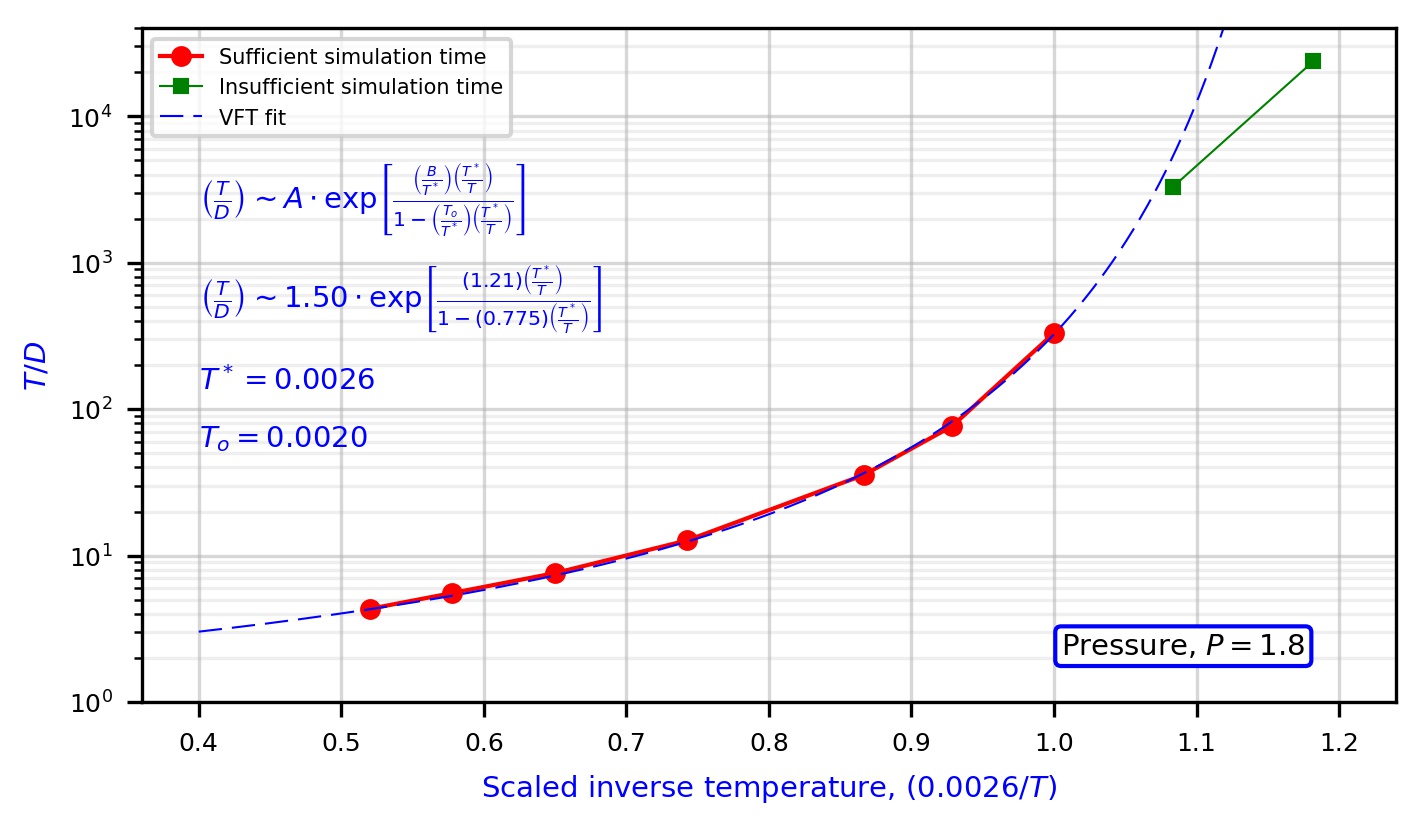}
\caption{
The dependence of the ``viscosity" (temperature-scaled inverse diffusion coefficient)
on the scaled inverse temperature $\sim (T_g/T) \approx (0.0026/T)$.
The results that are shown in the figure follow from the data presented in Fig. \ref{fig:meansquare}.
As we did not obtain reliable data for the behavior of the mean-square displacement at sufficiently
large times for temperatures $10^3 T < 2.6$ we assume that $10^3 T_g \sim 2.6$.
}\label{fig:etavsTgT}
\end{center}
\end{figure}

According to the Stokes-Einsten relation for the spherical particle of
radius $a$, the values of viscosity, $\eta$, and diffusion coefficient, $D$, are related via;
$\eta = (k_B/(6 \pi a))[T/D]$.
Therefore, under assumption that the Stokes-Einstein relation holds,
the ratio $T/D$ is commonly used to represent the results for the diffusion coefficient.
Thus, we plotted in Fig. \ref{fig:etavsTgT} the dependence of the ratio $T/D$ on
on the scaled inverse temperature, $(T^*/T)$.
Further, we fitted this dependence with the Vogel-Fulcher-Tammann (VFT) curve:
$\eta = A\exp\left[B/(T-T_o)\right]$.
The parameters of the fit are shown in the figure.

It is easy to estimate from the simulation data the analog of the fragility index, $m$ 
(``the analog" because we do not have the data at the true $T_g$). 
Thus, at $T=T^*=0.0026$ we get: $m = [\log(284.283)-\log(76.487)]/(1-0.0026/0.0028) = 0.57050/0.07143 = 7.99$.
This is a very small value and it follows from it that we are very far from $T_g$.

\begin{figure}
\begin{center}
\includegraphics[angle=0,width=3.4in]{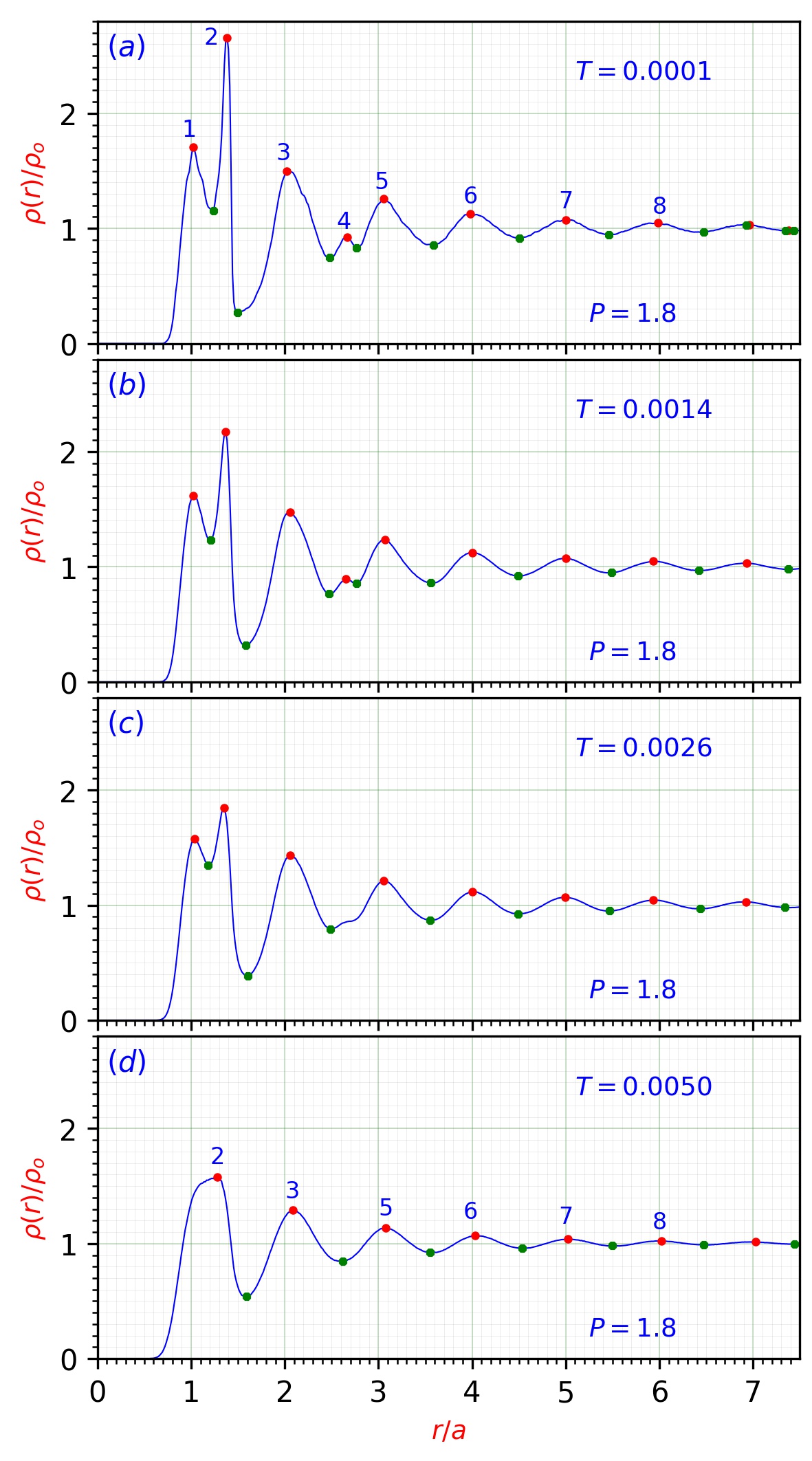}
\caption{
The dependencies of the scaled pair density functions, $g(r) \equiv [\rho(r)/\rho_o]$, on the
scaled distance, $[r/a]$, at the selected temperatures at $P=1.8$.
The scaling distance, $a$, is determined through the average density of $\rho_o \equiv 1/a^3$.
The order of the maximums and minimums, in the following discussions,
is determined by their order at $T=0.0001$, as follows from panels (a,d).
Note that the first peak, i.e., the pre-peak of the first major peak
is present only at sufficiently low temperatures, as well as the small forth peak.
}\label{fig:rhora-vs-ra}
\end{center}
\end{figure}

\begin{figure}
\begin{center}
\includegraphics[angle=0,width=3.5in]{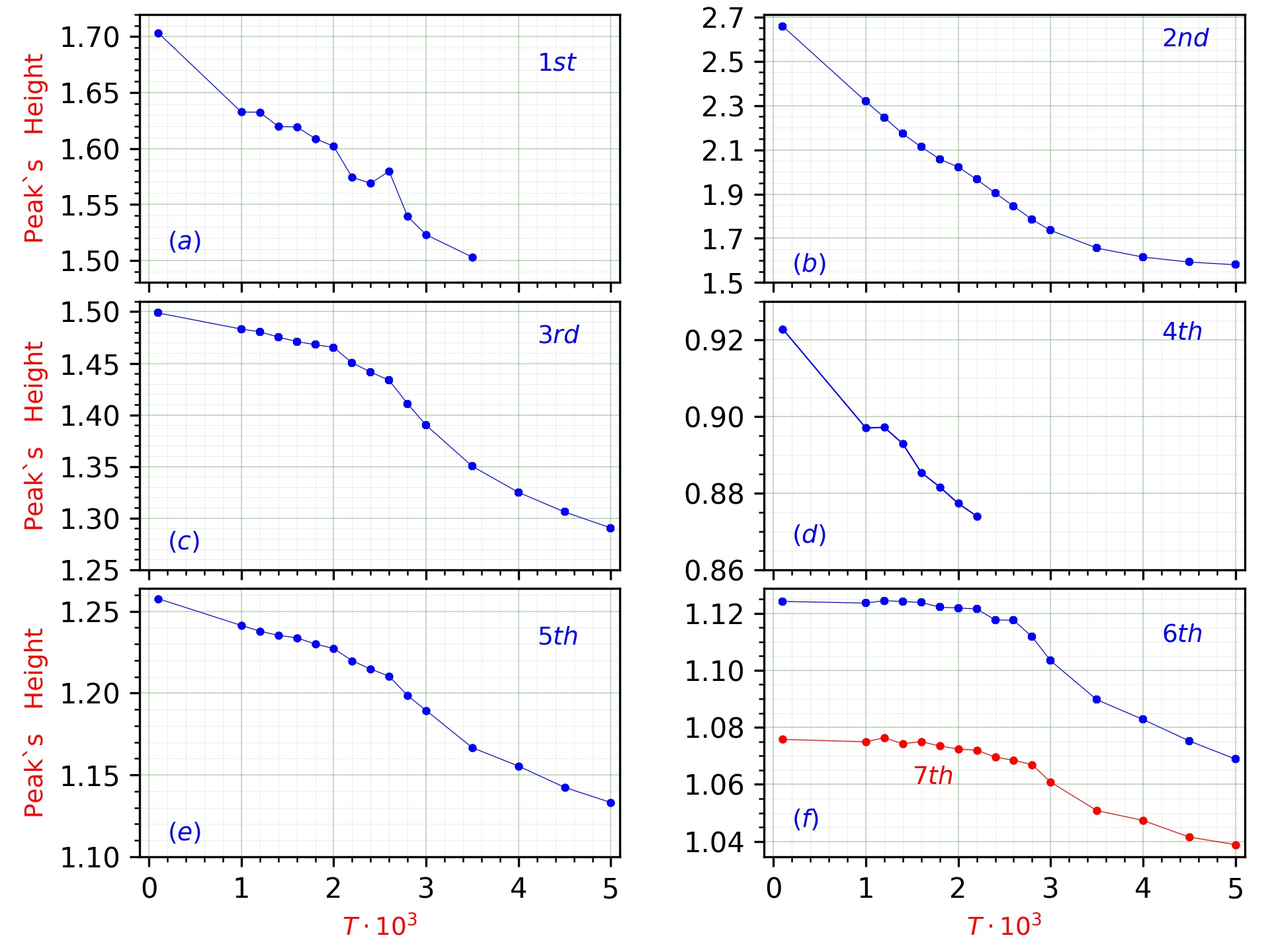}
\caption{
The dependencies of the heights of the peaks in the scaled PDF on the temperature.
The orders of the maximums/peaks are shown in panels (a,d) of Fig. \ref{fig:rhora-vs-ra}.
}\label{fig:rhor-max-min}
\end{center}
\end{figure}

\subsection{Results for the pair density function}

We now turn our attention to the description of the structural properties of the system.
As follows from the multiple previous considerations of the TCF and from section \ref{sec:triple},
it is reasonable to discuss the behavior of the TCF in connection with the behavior of the PDF.

Thus, in Fig.~\ref{fig:rhora-vs-ra} we show the PDFs calculated from the structures at the selected temperatures.
A noticeable feature of the shown PDFs is the development of the pre-peak of the 1st peak on the reduction of temperature.
While this feature is unusual for the systems with strong repulsion at short distances, it is well known for the soft and ultrasoft systems.
Another feature of interest is the development of a small peak at $r \approx 2.65$ as the temperature decreases.
As we already discussed in Ref.~\cite{levashov2019anomalous}, this feature resembles the splitting
of the second PDF peak, which is universally observed in supercooled liquids and glasses of different nature~\cite{Liu2010PRL,Ryltsev2013,Huerta2006PRE,Pan2011PRB,Trady2017JNCS,Mauro2011PRB,Wu2015NatureComm}. 
Such splitting is usually explained by the development of some complicated medium-range 
order as the system approaches glass transition~\cite{Pan2011PRB,Mauro2011PRB,Trady2017JNCS,Wu2015NatureComm}. 
This feature is one of the reasons that caused our interest in the further study of the structure
of the discussed system.

The proper integration of the first pre-peak shows that it is associated with approximately 6 or 7 particles.
Thus, it is logical to assume that ``an average" particle is at the center of the simple cube.
Therefore, the local structure might be similar to the {\it bcc} lattice.  Further integration of
the major part of the first peak shows that it is associated with approximately 12 particles.
Thus, it is logical to assume that these 12 neighbors approximately form a cuboctahedron, i.e.,
their positions are determined by the 180 degrees rotation of the central particle around the edges of the BCC cube.
Another alternative is that these 12 second neighbors form an icosahedron, though it is not clear how this geometry
can be consistent with the BCC neighborhood formed by the 1st neighbors.
In any case, both of these assumptions are not supported by our previous considerations of the BOOP. 
Moreover, the direct visual analysis of the nearest neighbor configurations for different 
particles shows such a significant amount of disorder that it is essentially impossible, 
in our view, to draw any definite conclusions about the nearest neighbor structure.

To further quantify the behavior of the PDFs in Fig.~\ref{fig:rhora-vs-ra}  we determined the positions and
the amplitudes of the maximums and minimums of these functions using the criteria for the local maximums and minimums.
These maximums and minimums are shown with the red and green symbols in Fig.~\ref{fig:rhora-vs-ra}.
Then we plotted the values of the PDFs at the maximums as the functions of the temperature,
as shown in Fig.~\ref{fig:rhor-max-min}.
It is of interest that the curves corresponding to the 3rd, 5th, 6th, and 7th maximums
exhibit changes in the slopes at temperatures which are close to $T=0.0026$,
which we consider to be the equilibrium breakdown temperature in our simulations.
On the other hand, the curves corresponding to the 1st and 2nd peaks
are not sensitive to this temperature.

\begin{figure*}
\begin{center}
\includegraphics[angle=0,width=6.5in]{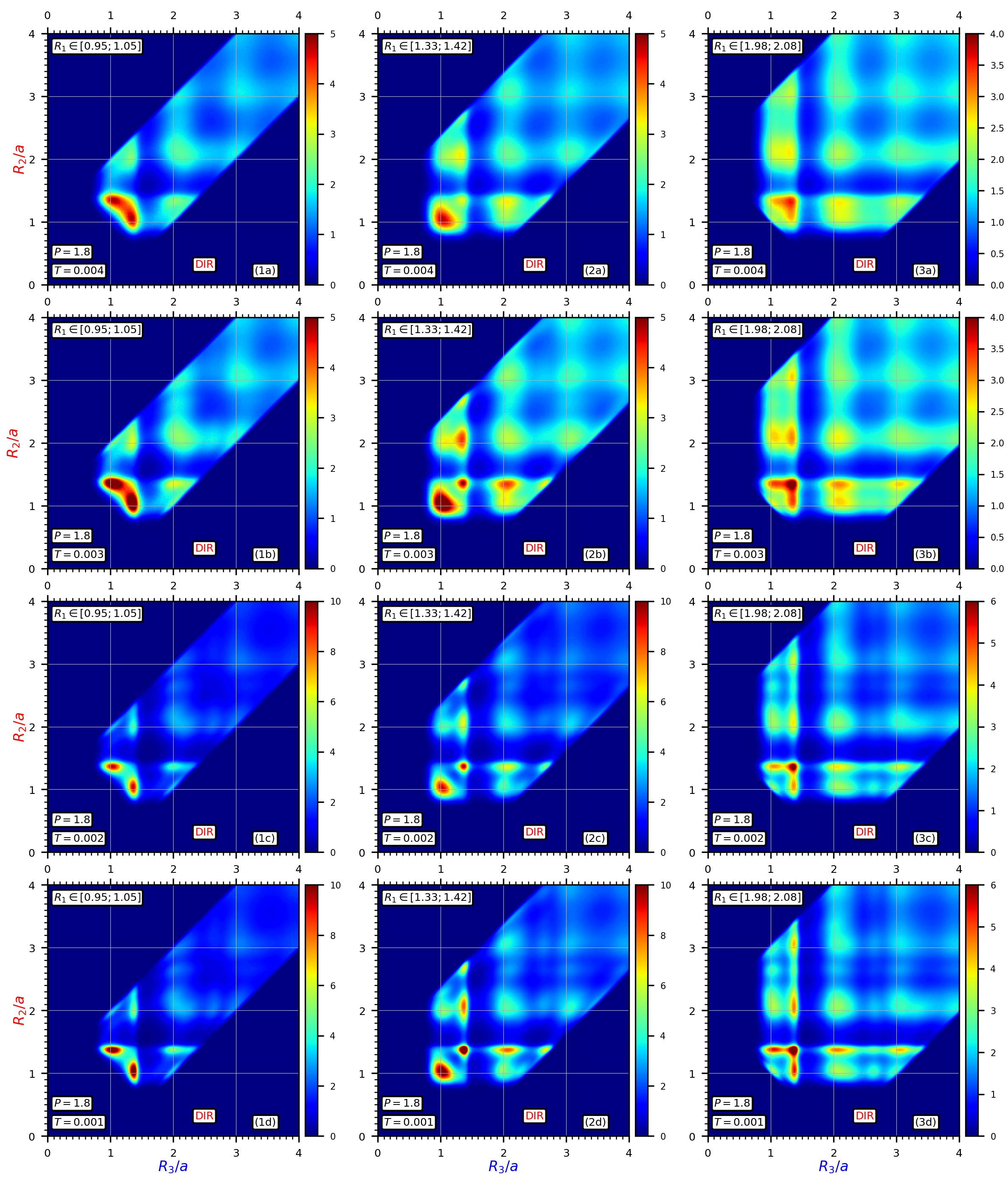}
\caption{
The 1st, 2nd, and 3rd columns show the evolution with the temperature of the TCF 
for the cases when the lengths of one side of
the considered triangles lie within 
distance intervals $[0.95,~1.05]$, $[1.33,~1.42]$, and $[1.98,~2.08]$ correspondingly.
Note that for the clarity of the data presentation different ``color-scales" were used for 
higher (panels 1a, 1b, 2a, 2b, 3a, 3b) and lower (panels 1c, 1d, 2c, 2d, 3c, 3d) temperatures.
Panels (1a, 1b, 2a, 2b, 3a, 3b) show the evolution of the TCF in the liquid state, i.e., increase in the peaks'
intensities associated with the decrease in the particles' mobilities and the decrease in their vibrational motion. 
Panels (1c, 1d, 2c, 2d, 3c, 3d) show the evolution of the TCF in the glass states, 
where the increase in the peaks' intensities is associated with the decrease in the vibrational motion only.
}\label{fig:TCF-2D}
\end{center}
\end{figure*}

\section{Results for the TCF}

If the PDF of the system at some conditions, $(T,P)$, is known, then it is possible to choose and approximately fix
the length of one side of the considered triangles at some value of interest, i.e., to fix approximately
one of the arguments of the TCF, and then consider how the TCF depends on the two remaining arguments.
Effectively, that is how the TCF is usually considered \cite{raveche1972three,bhatia1976triplet,haymet1981triplet,
haymet1985orientational,stillinger1988theoretical,muller1993triplet,ZahnK2003,Ru2003,ho2006three}.
Using this approach, we show in Fig.~\ref{fig:TCF-2D}
the results for the TCFs calculated for the triangles with the lengths of one side approximately fixed.
In particular, the first column of Fig.~\ref{fig:TCF-2D} corresponds to the case when
the lengths of one side of the chosen triangles are in the interval $[0.95,~1.05]$ which, 
according to Fig.~\ref{fig:rhora-vs-ra}, approximately corresponds to the position of the pre-peak of the first peak.
Then, the second column of Fig.~\ref{fig:TCF-2D} corresponds to the triangles with the lengths of one side 
in the interval $[1.33,~1.42]$.
This interval of distances, according to Fig.~\ref{fig:rhora-vs-ra},
approximately corresponds to the position of the major peak in the pair-density function.
Finally, the third column of Fig.~\ref{fig:TCF-2D} corresponds to the triangles with the lengths of one side 
in the interval $[1.98,~2.08]$. This interval of distances, according to panels (a,d) of Fig.~\ref{fig:rhora-vs-ra},
approximately corresponds to the position of the third peak in the pair-density function.

It follows from the 1st column of Fig.~\ref{fig:TCF-2D} that the smallest triangles in the system are not the equilateral triangles.
Indeed, for the chosen length of one side $\approx 1$ there are no triangles with two other sides equal to $\approx 1$.
Instead, as it follows from the 1st column, for the chosen side of length $\approx 1$ there are
triangles whose 2nd side is also $\approx 1$, while the 3rd remaining side has the length in the interval $\approx [1.30,~1.45]$.
Therefore, the smallest triangles in the system are the isosceles triangles.
We note that this information can not be extracted from the PDF in a simple and intuitive way.
Indeed, it follows from the KSA approach that every peak in the pair density function leads to the appearance
of the corresponding equilateral triangles.
It is known that this is not the case \cite{alder1964triplet,stillinger1988theoretical}.
Notice also that the system's organization does not favor the isosceles triangles with the smallest side $\approx 1.05$ and two other
sides of length $\approx 1.39$, as very clear from the (1d) and (2d) panels of Fig.~\ref{fig:TCF-2D}. 

The 2nd column of Figure~\ref{fig:TCF-2D} shows once again that the ``smallest" triangles (triangles with the smallest perimeter)  
in the system are the isosceles triangles whose largest side has length $\approx [1.30,~1.45]$,
while two smaller sides have lengths $\approx 1$.
It also follows from Fig.~\ref{fig:TCF-2D} that in the system there are equilateral triangles
with the sides of lengths $\approx [1.30,~1.45]$. 
These triangles correspond to the second peaks on the diagonals
that go from the bottom left to the top right corners of the 2nd column of Figure~\ref{fig:TCF-2D}.
The other triangles relatively well expressed in Fig.~\ref{fig:TCF-2D}
are the isosceles triangles whose largest side has length $\approx 2.10$,
while two smaller sides have lengths $\approx [1.30,~1.45]$.

The 3rd column of Fig.~\ref{fig:TCF-2D} shows the triangles whose one side has length in the interval  $\approx [1.98, 2.08]$.
It follows from the third column that there are the following relatively well expressed triangles in the system:
$ \approx $ $[1.35,~1.35,~2.03]$, $[1.05,~1.35,~2.03]$, $[1.35,~2.03,~2.03]$, $[1.35,~2.03,~3.05]$.
Some other less well expressed triangles are 
$ \approx $ $[1.05,~1.05,~2.03]$, $[1.05,~2.03,~2.15]$, $[2.03,~2.03,~2.03]$, $[1.05,~2.03,~2.70]$.

An important point to notice concerns a natural limitation of the considered representation for the TCF.
Thus, note that from the 1st column of Fig.~\ref{fig:TCF-2D} alone we can not make a conclusion about the presence
or absence in the system of the equilateral triangles with the sides of length $\approx [1.30, 1.45]$.
To gain this information we need to consider the 2nd column of Fig.~\ref{fig:TCF-2D}, i.e., the TCF with
one parameter fixed on the distance of interest $\approx [1.30, 1.45]$. 
Similarly, from the 2nd column of Fig.~\ref{fig:TCF-2D} alone we can not conclude if there are in the system 
the equilateral triangles with the sides of length $[0.95,~ 1.05]$.
For this it is necessary to consider the TCF with one argument fixed at $\approx [0.95,~ 1.05]$, i.e., 
the 1st column of Fig.~\ref{fig:TCF-2D}.
Despite these limitation, in our view, Fig.~\ref{fig:TCF-2D} provides valuable
intuitive insights into the strengths of the angular correlations in the studied liquid.
Note that Fig.~\ref{fig:TCF-2D}, in general and especially its 3rd column, addresses also
the angular correlations at distances beyond the nearest neighbors. 

It is clear from the previous considerations that it is possible to extract
from the TCF information about the presence or absence in the system of
particular equilateral or isosceles triangles, i.e., the information
which is not possible to gain from the PDF simply and intuitively.
Since the TCF allows estimating the strength of the angular orientations in liquids,
it is of interest to investigate this issue in a more quantitative way.
In particular, these considerations are relevant since our previous investigations,
in terms of the BOO parameters, were difficult to interpret.

\begin{figure}
\begin{center}
\includegraphics[angle=0,width=3.5in]{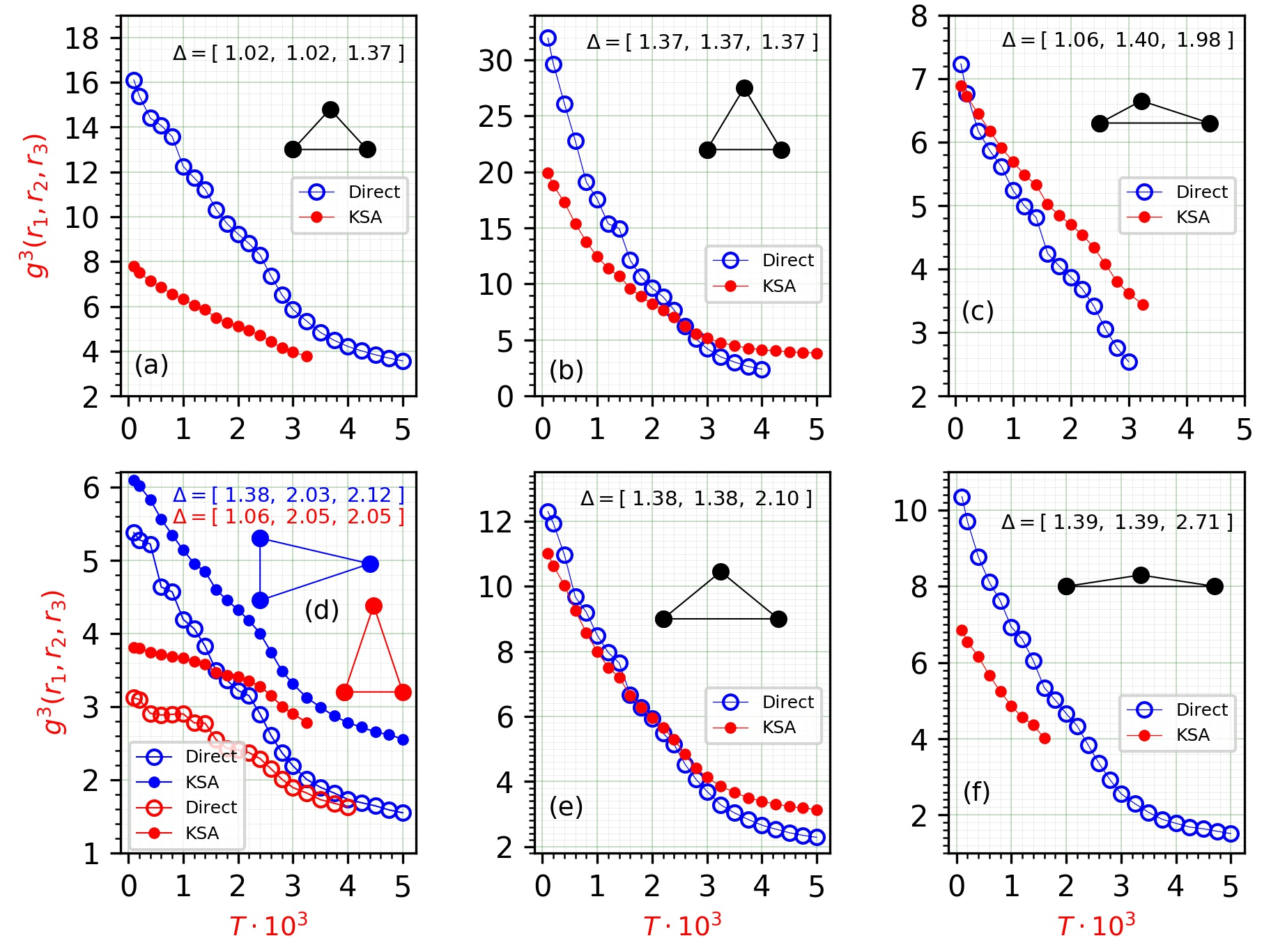}
\caption{
The dependencies of the peaks' heights of the triple correlation function on temperature.
The considered triangles are the triangles with the relatively small side lengths.
For example, the curve in panel (a) corresponds to the triangle with the smallest perimeter,
which happens to be an isosceles triangle. Two equal sides of this triangle correspond to the distance between the nearest
neighbors, while another side corresponds to the distance between the second neighbors.
Note that the smallest triangle is not an equilateral triangle.
It follows from the figure that considerations of the maximums in the TCF provide a simple
and intuitive way to address the average strength of the angular correlations in liquids.
}\label{fig:triplepeaks}
\end{center}
\end{figure}

Figures \ref{fig:rhora-vs-ra},\ref{fig:rhor-max-min} address the evolution with temperature of the PDF.
It is possible to perform a similar analysis with the TCF. For this it is necessary 
at first to produce a 3D array of the TCF, i.e., an array that
shows the number of triangles with the sides of length $(r_1,r_2,r_3)$.
Further, the local maximums of the TCF can be found in a straightforward way, i.e.,
some local maximum should be the absolute maximum within some range of values $(r_1 \pm dr_1)$ and similar for the $r_2$ and $r_3$.
We note that this type of analysis, as well as the calculation of the TCF, can be performed quite quickly nowadays even on a single processor.
In this way, we determined the locations and the magnitudes of the maximums by analysis of the lowest temperature structures.
In consideration of the maximums from the higher temperatures, we were associating a maximum from a higher-temperature structure
with a maximum from the lowest-temperature structure if ``the locations'' $(r_1,r_2,r_3)$
of these maximums were within some allowable range, $\Delta$, from each other for each argument, i.e., $r_1$, $r_2$, and $r_3$ separately.
In our considerations, we used the range value of $\Delta =0.1 a$.
This approach allowed us to address the evolution of the maximums quite well.
The obtained results are shown in Fig.~\ref{fig:triplepeaks}.
In Fig.~\ref{fig:triplepeaks} we also show with the red symbols and curves the evolution
of the TCF maximums for the selected (truly existing) triangles according to the KSA (\ref{eq:ksa1}).
In the analysis of the data shown Fig. \ref{fig:triplepeaks}
it is very useful to check the locations of the considered triangles
in Fig.~\ref{fig:TCF-2D}.

In panel (a) of Fig.~\ref{fig:triplepeaks} we address the evolution with the temperature of the maximum
of the TCF corresponding to the smallest (isosceles) triangle in the system. It follows from the blue curve
that the magnitude of the maximum, as the temperature is reduced from $10^3 T = 5.0$ to $10^3 T = 0.1$,
increases $\approx 4.5$ times. The comparison of the blue and red curves shows that the KSA
fails to describe adequately the actual situation. The comparison of the results presented
in this panel with the data presented in Fig. 15
of Ref.~\cite{levashov2019anomalous} shows that the considerations of the TCF indeed provide additional
insight into the strength of the angular correlations. It is arguable if the blue curve exhibits
some change in behavior at the temperature corresponding to the equilibrium breakdown
temperature in our simulations, i.e., at $10^3 T \approx 2.6$.
There is an additional interesting point concerning the results in panel (a).
Note that the KSA curve ends at temperature $10^3 T = 3.5$, i.e., at the temperature
when the pre-peak of the first peak
in PDF disappears. However, the analysis of the TCF calculated directly still shows
the existence of the maximum corresponding to the considered isosceles triangle.
This difference shows that the TCF, in comparison to the PDF, is more sensitive to the structural changes.

We note here again that the direct visual analysis of the atomic environments 
does not provide an understanding with respect to the local atomic structure at the considered pressure.

Panel (b) of Fig.~\ref{fig:triplepeaks} addresses the evolution with the temperature 
of the maximum of the TCF corresponding to the equilateral triangles formed by the second neighbors
(see 2nd column of Fig.~\ref{fig:TCF-2D}).
We see that the KSA curve is quite close to the curve calculated directly from the structures at high temperatures.
However, as the temperature is reduced, the disagreement becomes more significant and at the lowest temperature 
the KSA approach underestimates the strength of the triple angular correlations quite significantly.

With respect to panels (c,d,e), it is clear the KSA starts to perform better for the triangles with sides
larger than the smallest interparticle separation distance.
All panels suggest a significant increase of the angular correlations which
is not obvious from the behavior of the BOOPs shown in Fig.15 of Ref.~\cite{levashov2019anomalous}.

Panel (f) is of special interest.
One of the sides of the corresponding triangle has length $\approx 2.71$ which, according to panel (a) of Fig.~\ref{fig:rhora-vs-ra},
corresponds to the small 4th peak which develops in the low-temperature glassy state and which is absent in the liquid state.
 That is why the KSA curve ends at $10^3 T = 1.4$.
On the other hand, like with panel (a) of Fig.~\ref{fig:triplepeaks},
we see that analysis of the TCF calculated directly from the structures still finds the
corresponding maximum. Also, note that there is a very significant disagreement between
the magnitude of the maximum calculated directly from the structures and the magnitude of the peak
obtained from the KSA.
Further, note that the considered triangle essentially is formed by the nearly aligned particles.
Indeed, $1.39 + 1.39 = 2.78 \approx 2.71$.
Thus, development of the alignment of the central chosen particle and its two second neighbors
appears to be connected to the development of the 4th peak in the PDF, as shown in Fig.~\ref{fig:rhora-vs-ra}(a,b).
Note that this information can not be easily extracted from the behavior of
the PDF via the KSA. This information also does not easily follow from the analysis of the BOOP.
Finally, note that we consider the angular ordering of the 2nd and
not the 1st neighbors in Fig.~\ref{fig:triplepeaks}(f).

\section{Investigations on the peaks' shapes}

In this section, we describe the results of some of our investigations concerning 
the evolutions of the peaks' shapes with the temperature. 
The principal purpose of these considerations was to check if it is possible to see in the behavior of 
the average TCF the presence of the glass transition. In any case, these studies also 
provide additional insight into the features of the TCF.

A particular technical point that is relevant to the present considerations 
is related to our definition of the TCF in Eq.~\ref{eq:tcfdef}. 
According to this definition, in the case of the completely random 
distribution of particles, the introduced TCF is equal to one. 
In our view, it is reasonable to redefine the TCF in such a way that 
it would be equal to zero for the completely random, i.e., uncorrelated case. 
Thus, we find it reasonable, in defining the TCF 
for this section, to subtract one from the definition (\ref{eq:tcfdef}).

\begin{figure*}
\begin{center}
\includegraphics[angle=0,width=7.0in]{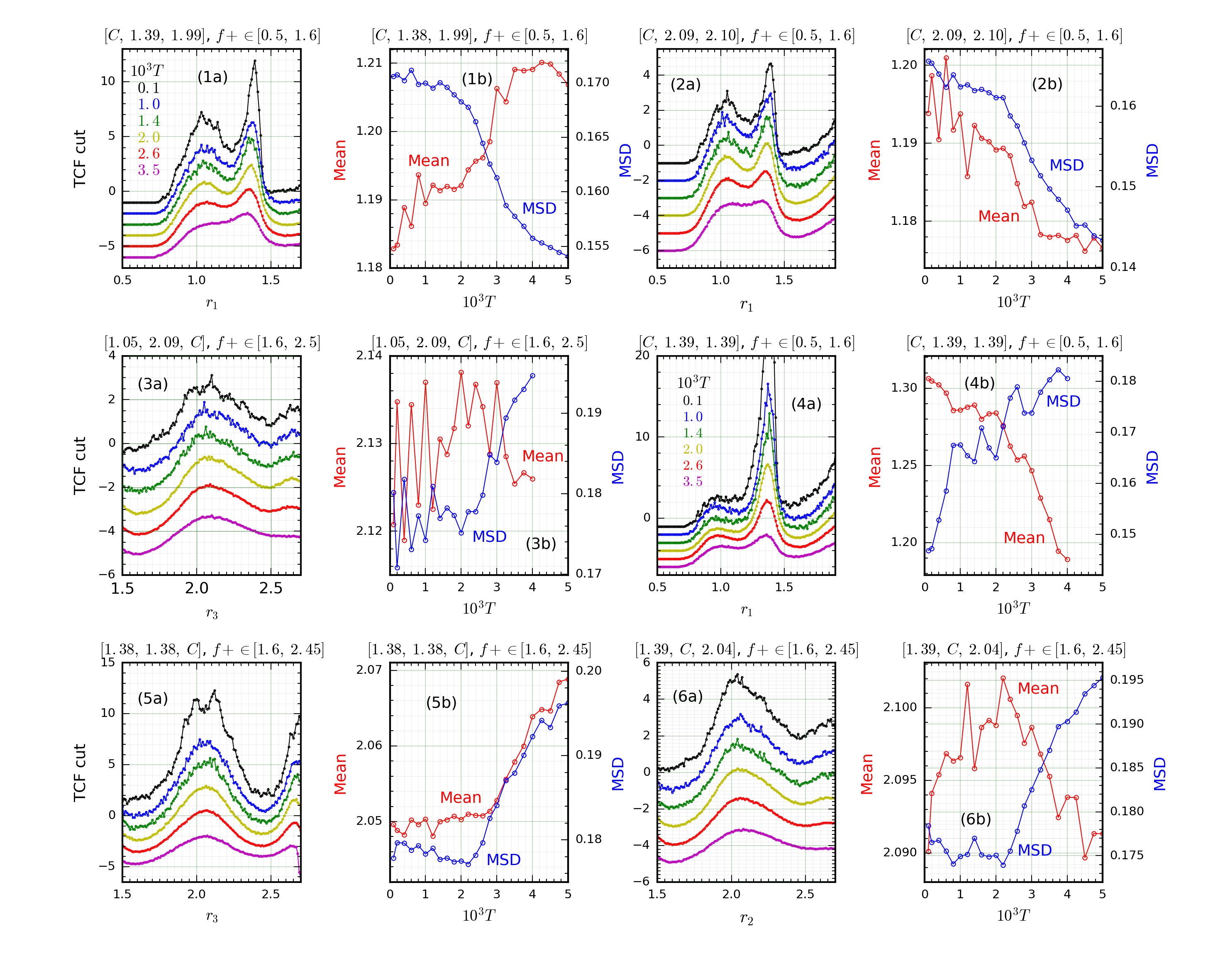}
\caption{
Panels (1a, 2a, 3a, 4a, 5a, 6a) show the selected cuts of the TCF functions when one of its arguments changes, 
while the two other arguments fixed at the values corresponding to the ``position" of the chosen peak in the TCF.
Panels (1b, 2b, 3b, 4b, 5b, 6b) show the dependencies on the temperature of the mean values and 
the mean square deviations for the cuts shown in the ``a"-panels. 
The values of the means and the MSDs were calculated from the segments of 
the curves located between the appropriate minimums, as described in more detail in the text.
}\label{fig:widths1}
\end{center}
\end{figure*}

In panel (1a) of Fig.~\ref{fig:widths1} we show the cuts 
of the redefined TCF in the vicinity of the peak associated with 
the triangles with the sides' of lengths $[\approx 1.02,\;\approx 1.39,\;\approx 1.99]$. 
We show the curve corresponding to the lowest temperature as it is.
Then, for the clarity of the presentation, we shifted the curves 
corresponding to higher temperatures downwards by $1.0$, $2.0$, e.t.c.
In the title of subplot (1a) and similar other subplots 
of Fig.~\ref{fig:widths1} the letter ``C" stands for the ``curves" shown in the subplots.
Thus, in subplot (1a) we show curves corresponding to the triangles
with the two fixed sides ($1.39$ and $1.99$), while the length of the 3rd side varies.
The distance resolution of our TCF is $0.01$.
It is convenient and very useful to identify with the curves shown in 
Fig. \ref{fig:widths1}(1a) the corresponding intensities 
in the 2nd and 3rd columns of Fig.~\ref{fig:TCF-2D}.

Another technical comment is related to how we obtained the shown curves at higher temperatures.
The procedure was similar to the one already described in the context of Fig.~\ref{fig:triplepeaks}.
Thus, at every higher temperature, we determined ``the coordinates" of the local maximum corresponding 
to the chosen peak at the lowest studied temperature. The criterion for the peaks' correspondence 
was the same as for Fig.~\ref{fig:triplepeaks}, i.e., peaks from the two different temperatures 
were classified as the same if their coordinates ``along all three directions" were within $0.10a$. 
Then, the cuts along $r_1$ or $r_2$, or $r_3$ of the TCF we performed through the found coordinate of the maximum. 
Thus obtained cuts are shown in Fig.~\ref{fig:widths1}(1a,2a,3a,4a,5a,6a).

In principle, one may assume that the merging of the two peaks in the curves in Fig.~\ref{fig:widths1}(1a) 
is related to the happening transitions between the triangles 
$\left\{\approx [1.02,\;1.39,\;1.99]\right\} \leftrightarrow \left\{\approx [1.39,\;1.39,\;1.99]\right\}$.
See also the 2nd and the 3rd columns in Fig.~\ref{fig:TCF-2D} in this context.
However, the results for the mean square displacements shown in Fig.~\ref{fig:meansquare} do not support this view.
Indeed, the position of the plateau in Fig.~\ref{fig:meansquare} at low temperatures suggests that, 
as the particles vibrate near their equilibrium positions, they deviate from these positions by $\sim 0.1a$.
This distance is approximately three or four times smaller than the separation between the peaks' maximums.
Thus, we have to conclude that the spread of both peaks and their overlaps are mostly caused by 
a static structural disorder. This result is expectable in the context of the multiple previous investigations of the other systems.

Further analysis of the peaks' shapes is complicated because of the complex shapes of the shown curves. 
For example, we found that it is impossible to fit well the shown curves with the two Gaussian functions.
Moreover, there is no reason to expect that it should be possible to fit these curves with any other simple functions.
Therefore, using the fact that the curves in Fig.~\ref{fig:widths1}(1a) contain some relatively well-defined minimums, 
we calculated the average values and the mean square deviations (MSDs) of the curves' segments 
located between the two selected minimums.

Thus, we chose the segment associated with the peaks shown in panel (1a) of Fig.~\ref{fig:widths1} to be $[0.5,\; 1.6]$. 
This choice of the segment is reflected in the titles of panels (1a) and (1b). 
As we discussed previously in this section, it is reasonable to associate 
with the positive correlations the positive values of the ``redefined" TCF. 
Thus, in our view, it reasonable, in calculating the average values and the MSDs 
of the curves shown in Fig.~\ref{fig:widths1}(1a) to take into account only the positive values of the redefined TCF. 
This choice of only the positive values of the redefined TCF in 
the selected segments is reflected in the titles of the plots by the notation ``$f+ \in$".

In panel (1b) of Fig.~\ref{fig:widths1} we show the dependencies on the temperature of the average 
value and the MSD of the chosen segments of the curves shown in panel (1a) of Fig.~\ref{fig:widths1}. 
The left axis corresponds to the mean value (red curve), while the right axis corresponds to the MSD (blue curve).
The other panels of Fig.~\ref{fig:widths1} organized in a way that is identical 
to the organization of the just described panels (1a) and (1b). 
In panels (2a, 4a, 6a) we omitted the notation ``TCF cut" on the $y$-axis for the compactness of the layout.

Note in panels (1b, 2b, 3b, 4b, 5b, 6b) of Fig.~\ref{fig:widths1} that certain curves for 
the dependence of the mean on the temperature (1b, 5b, 6b)
and especially for the dependence of the MSD on the temperature (1b, 2b, 3b, 5b, 6b) appear 
to be systematically sensitive to the glass transition temperature. 
Especially, note that the most pronounced changes in the behavior of the MSD curves happen
for the single peaks shown in panels (3a,5a,6a) with the corresponding MSD curves in panels
(3b,5b,6b). We stress attention on this point because for the single peaks the analysis that we perform
is less questionable than for the double peaks.
However, note also that the considered changes in the means and MSDs, as the temperature changes in the whole
studied range are quite small. Thus, for the MSDs in panels (3b,5b,6b) this change is approximately $10\%$.

In our view, before making more definite conclusions concerning the presented results 
it is reasonable to consider the behavior of the described parameters in the other glass-forming
systems.

\section{Conclusion}\label{sec:conclusion}

The main purpose of this paper was to investigate with the triple correlation function (TCF) the development of 
the orientational ordering under supercooling in a simple ultrasoft fluid.
The considered model fluid consisted of particles interacting through the harmonic-repulsive pair potential. 
We studied the system at reduced pressure P = 1.8 at which it exhibits remarkable stability against crystallization. 
The choice of the TCF, as the method of investigation, is caused by the results of our previous studies with 
the pair distribution function (PDF) and the bond-orientational order parameters (BOOPs). 
These methods, as well as the visual analysis of the structures, did not provide satisfactory insights 
into the organizations of the local atomic environments. 
Therefore, we attempted to address the strength of the orientational correlations 
in the simplest structural units, besides the particles' pairs.

The obtained results clearly show the development of the orientational correlations on cooling.  
In particular, it follows from the data that naive expectations, based on 
the Kirkwood's superposition approximation, significantly underestimate 
the strength of the orientational correlations for certain triangles. 
These underestimates are the most significant for the triangles with 
the side lengths corresponding to the 1st and 2nd neighbors. 
However, a significant underestimate happens also for the triangle 
associated with the splitting of the second peak in the PDF, which, 
as it follows from the results, corresponds to the alignment 
of the central chosen particle with its two 2nd neighbors.

Detailed considerations of the peaks' shapes in the TCF suggest that
there exists a connection between the orientational ordering and the
slowdown of the system's dynamics. Because of the importance of this result, 
this issue requires further study.

Although, in general, the method of the TCF is well known, 
it is relatively rarely used in the analysis of the liquid structures. 
In our view, this method should be used more often in the analysis 
of the simulated data because it provides additional significant insights 
into the structural organizations of the considered systems. 
Especially, these considerations can valuably supplement studies with the BOOPs.

\section{Acknowledgements} 

This work was supported by the Russian Science Foundation (grant 18-12-00438). 
We gratefully acknowledge access to the following computational resources: 
Supercomputing Center of Novosibirsk State University (http://nusc.nsu.ru), 
the federal collective usage center ‘Complex for Simulation and Data Processing 
for Mega-science Facilities’ at NRC ‘Kurchatov Institute’ (http://ckp.nrcki.ru/), 
supercomputers at Joint Supercomputer Center of Russian Academy 
of Sciences (http://www.jscc.ru), and ‘Uran’ supercomputer 
of IMM UB RAS (http://parallel.uran.ru).




\bibliographystyle{zbib-unsrtnat}

\bibliography{tcf-full-2020-05-19-1}

\end{document}